\documentclass[twocolumn,preprintnumbers,amsmath,amssymb,superscriptaddress]{revtex4}
\UseRawInputEncoding
\usepackage{graphicx}
\usepackage{dcolumn}
\usepackage{bm}
\usepackage{soul}
\usepackage{color}
\usepackage{epstopdf}
\usepackage[version=3]{mhchem}
\usepackage{lipsum}
\usepackage[outercaption]{sidecap}
\usepackage{floatrow}
\begin{document}


\title[Sample title]{Screening and collective effects in randomly pinned fluids: A new theoretical framework}

\author{Anh D. Phan}
\affiliation{Faculty of Materials Science and Engineering, Phenikaa Institute for Advanced Study, Phenikaa University, Hanoi 12116, Vietnam}

\email{anh.phanduc@phenikaa-uni.edu.vn}

\date{\today}

\begin{abstract}
We propose a theoretical framework for the dynamics of bulk isotropic hard-sphere systems in the presence of randomly pinned particles and apply this theory to supercooled water to validate it.  Structural relaxation is mainly governed by local and non-local activated process. As the pinned fraction grows, a local caging constraint becomes stronger and the long range collective aspect of relaxation is screened by immobile obstacles. Different responses of the local and cooperative motions results in subtle predictions for how the alpha relaxation time varies with pinning and density. Our theoretical analysis for the relaxation time of water with pinned molecules quantitatively well describe previous simulations. In addition, the thermal dependence of relaxation for unpinned bulk water is also consistent with prior computational and experimental data.
\end{abstract} 

\maketitle

\section{Introdction}
After several decades of intensively attempting to understand fully physical mechanisms underlying glass transition, this phenomenon has still challenged scientists all over the world as an unsolved and fascinating problem \cite{1,2,3}. When molten materials are cooled down by fast rate to avoid crystallization, the systems fall out of equilibrium. The structural relaxation time of these non-equilibrium substances dramatically increases many orders of magnitudes and exceeds both simulation and experimental time scales. While recent breakthrough developments in glass theory enable to access low temperatures far from the glass transition temperature, $T_g$, and analyze roles of decisive factors responsible for kinetic slowdown. Understanding how to manipulate the glass transition would open revolution in a wide range of technological applications and fundamental physics. 

It is well-known that spatial confinement effects substantially vary glassy dynamics of a system compared to its bulk counterpart \cite{4,5}. There are many ways to design confinement: pinning molecules and/or particles \cite{6,7,8,29,30,31}, quenching network \cite{9} or using surfaces and interfaces \cite{10,11,12,13}. Dynamics of confined systems can be slower \cite{10,11,12,13} or faster \cite{14,15,16} since molecular mobility strongly depends on properties of surfaces, boundaries and finite-size effects. Although much effort has been dedicated to obtaining a universal description for the dynamical property, it is still poorly understood and a debate topic. Among such confinements, influences of random pinning process on the structural relaxation of glass formers have received particular attention due to imitability of quenched disorder porous media and the slowing down dynamics of mobile particles without changing the structure.

Recently, Elastically collective nonlinear Langevin equation (ECNLE) theory has described qualitatively and quantitatively the alpha and beta relaxation, fragility, diffusion constant, and mechanical properties of various single-component materials, composites, and films  \cite{16,17,18,27,Phan2019,PhanPRL,35,40,41,42,21,22,24}. Basically, the theory views a relaxation event as a combination of two distinctive but strongly related processes: cage-scale hopping and the long range elastically collective rearrangement of surrounding particles. Local caging constraints play a main role on structural relaxation at high temperatures but the elastic deformation becomes dominant in deeply supercooled regime. The presence of confinement changes correlation between local and collective dynamics. Thus, The relaxation and other related properties are affected. Since the ECNLE theory predicts relaxation time in the range of 0.1 ps to $10^3$ s, it can be exploited to explain physical mechanisms underlying measurements in simulations and experiments. 

In a prior work \cite{27}, we extended the ECNLE theory to qualitatively elucidate trends seen in simulations of slow dynamics in randomly pinned fluids. Immobile particles occupy available space and restrict motion of mobile particles. The presence of pinned particles reduces the mobility of particle cage and greatly enhances local trapping effects. However, pinned particles introduce inhomogeneity of the fluid and are unresponsive to cooperative rearrangement. These features raise an idea of screening or spatial localization of the distortion/displacement field associated with cooperative motions beyond the local cage scale. However, for simplification, we completely ignored the screening effect on the glass transition in Ref. \cite{27}. The missing physics leads to a large quantitative deviation between theory and simulations. Thus, with screening effect of elasticity, key intriguing and open questions include: (1) how does the elastic displacement field behave? (2) how does a role of collective motions on the glass transition change with pinning? (3) can new theory provide a better description for simulations and experiments? Answers for these questions are not trivial, and touch on interesting fundamental physics issues.

In this work, we develop a new framework of the ECNLE theory to capture the screening effect caused by immobile components on the structural relaxation of randomly-pinned particle systems. After formulating a mathematical form of the screened displacement field, the longer range collective elastic part of the barrier is calculated. Contribution and competition of the local cage effects and collective rearrangement to the dynamical relaxation become subtle. This behavior totally different from the previous study \cite{27}. To evaluate validity of the new theory, we use this approach to model dynamics of supercooled water and compare results with previous simulation and experimental studies. 
\section{Theoretical Background}
We consider glassy dynamics of a single-component hard-sphere system but a finite fraction of particles are randomly pinned to provide neutral confinement. This pinning process has no significant effect on the structure. The idealized treatment is consistent with prior simulations \cite{7,8} which Vogel and his coworkers simulated a system of pinned water molecules and showed that the radial distribution functions, $g(r)$, are insensitive to pinning. The direct correlation function of mobile particles is identical to that of pinned particles $c_{11}(r)=c_{12}(r)=c_{22}(r)=c(r)$ \cite{27}, where $c(r)$ is the direct correlation function of the one-component hard sphere fluid, and the subscript 1 and 2 denote mobile and pinned particles, respectively. For a fluid of unpinned bulk hard spheres, the Percus-Yevick integral equation theory can be used to compute structural correlations. 

For sufficiently large density, a mobile particle is dynamically arrested within a particle cage formed by its nearest neighbors and a barrier emerges in the dynamic energy profile. The dynamic free energy of a tagged mobile particle in the pinned-mobile sphere systems quantifying its interactions with nearest neighbors is calculated by \cite{20,27}
\begin{eqnarray}
\frac{F_{dyn}(r)}{k_BT} &=& - \int \frac{d\mathbf{q}}{(2\pi)^3}\left[\frac{c(q)S_{12}(q)}{\rho(1-\alpha)\left[1-\rho(1-\alpha) c(q) \right]}\right.\nonumber\\
&+&\left.\frac{\rho(1-\alpha)c^2(q)e^{-q^2r^2\left[1-\rho(1-\alpha)c(q)\right]/6}}{\left[1-\rho(1-\alpha)c(q)\right] \left[2- \rho(1-\alpha)c(q)\right]} \right]e^{-q^2r^2/6}\nonumber\\
&-& 3\ln\left(\frac{r}{d}\right),
\label{eq:1}
\end{eqnarray}
where $r$ is the displacement of the particle, $d$ is the particle diameter, $q$ is the wavevector, $k_B$ is the Boltzmann constant, $T$ is the temperature, $\rho$ is the density number of a hard-sphere fluid or the number of hard spheres per volume, $\alpha$ is the pinning fraction, $c(q)$ is the Fourier transform of $c(r)$, and the static structure factor $S_{12}(q)$ is given by 
\begin{eqnarray}
S_{12}(q) &=& \frac{\sqrt{\rho_1\rho_2}c_{12}(q)}{[1-\rho_1c_{11}(q)][1-\rho_2c_{22}(q)]-\rho_1\rho_2c_{12}(q)c_{21}(q)}, \nonumber\\
\end{eqnarray}
where $\rho_1=\rho(1-\alpha)$ and $\rho_2=\rho\alpha$ are the density number of mobile and pinned particles, respectively. The leading term in Eq. (\ref{eq:1}) corresponds to a dynamic mean-field trapping potential favoring localization and the second term favors the fluid state. 

Numerical calculations of $F_{dyn}(r)$ provide information of characteristic length and energy scales of local dynamics as shown in Fig. \ref{fig:3}a. The local minimum and maximum of $F_{dyn}(r)$ are the localization length, $r_L$, and the barrier position, $r_B$, respectively. The local barrier height is $F_B = F_{dyn}(r_B)-F_{dyn}(r_L)$ and the jump distance from the localized position to the barrier position is $\Delta r = r_B-r_L$. 

When $\alpha = 0$, the dynamic free energy of two-component system reduces to that of one-component system
\begin{eqnarray}
\frac{F_{dyn}(r)}{k_BT}&=&-3\ln\left(\frac{r}{d}\right) -\int\frac{d\mathbf{q}}{(2\pi)^3}\frac{\rho c^2(q)}{\left[1-\rho c(q) \right]\left[2-\rho c(q) \right]}\nonumber\\
&\times&\exp\left[-\frac{q^2r^2}{6}\left(2-\rho c(q)\right) \right].
\end{eqnarray}

The pinning process induces slower particle dynamics due to a stronger local caging constraint. Apart from local dynamics, the structural relaxation is also governed by longer-range collective motions. Diffusion of a particle from its cage requires cooperative rearrangement of particles in the first coordination shell to create space for a large amplitude hop and displacement field, $u(r)$, outside the cage along radial direction. Physically, we expect that pinned particles act as obstacles which can "screen" the displacement field associated with the collective elastic component of the activation barrier. The long range displacements are sufficiently small, so we model the motion around their localized position as harmonic oscillations. The spring constant in ECNLE theory is calculated using $K_0\equiv K_0(\alpha)=\left.\cfrac{\partial^2 F_{dyn}(r)}{\partial r^2}\right|_{r=r_L}$. Pinned particles are made stationary by applying a hypothetical forces to balance the internal elastic force on the particle. The structure of pinned-mobile particle system is identical to that of a one-component system, and the Hookean restoring force acting on a mobile particle $K_0(\alpha)u(\alpha)$ is the same as the elastic force exerting on the pinned particles. Thus, the external force $\mathbf{f}_{ext}$ has to be $-K_0(\alpha)u(\alpha)$. The '-' sign illustrates the opposite direction of $\mathbf{f}_{ext}$ relative to the elastic force. Additionally, the equilibrium equations and boundary conditions for the continuum elastic treatment of collective barrier remain unchanged.

The continuum elastic equation for the displacement field beyond the cage scale is 
\begin{eqnarray}
\left(\mathcal{K}_B+\frac{G}{3}\right)\nabla(\nabla .\mathbf{u}) + G\nabla^2\mathbf{u} + \rho\alpha\mathbf{f}_{ext} = 0,
\label{eq:40}
\end{eqnarray}
where $\mathcal{K}_B$ and $G$ are the bulk and dynamic shear modulus, respectively, and $\rho\alpha\mathbf{f}_{ext}$ is the body force (or force per unit volume). Solving Eq. (\ref{eq:40}) with boundary conditions of $u(r=r_{cage})=\Delta r_{eff}$ and $u(r\rightarrow\infty)=0$ gives an analytical result of the displacement field
\begin{eqnarray}
u(r) = \frac{1 + \kappa r}{1 + \kappa r_{cage}}\frac{e^{-\kappa (r-r_{cage})}}{r^2}\Delta r_{eff}r_{cage}^2,
\label{eq:dis}
\end{eqnarray}
where $r_{cage}$ is the cage radius estimated by the first minimum position of $g(r)$ and an amplitude of the cage expansion, $\Delta r_{eff}$, is \cite{19,27}
\begin{eqnarray}
\Delta r_{eff} =\frac{3}{r_{cage}^3}\left[\frac{r_{cage}^2\Delta r^2}{32}-\frac{r_{cage}\Delta r^3}{192}+\frac{\Delta r^4}{3072} \right],
\label{eq:screen}
\end{eqnarray}
and $\kappa$ is the inverse screening length of the displacement field given by
\begin{eqnarray}
\kappa = \sqrt{\frac{6\alpha\Phi}{\pi d^3}\frac{K_0}{\mathcal{K}_B+4G/3}}, \end{eqnarray}
where $\Phi=\rho\pi d^3/6$ is the volume fraction and the bulk modulus of the hard-sphere fluid is \cite{19}
\begin{eqnarray}
\mathcal{K}_B = \frac{6\Phi k_BT}{\pi d^3}\frac{(1+2\Phi)^2}{(1-\Phi)^4}.
\label{eq:screen}
\end{eqnarray}
Based on the systematic analysis of the dynamic shear modulus in the ultra-local limit \cite{23}, $G$ can be approximated as $9\Phi k_BT(1-\alpha)/5\pi dr_L^2$. The displacement field is screened in the same manner as the Yukawa-like function.

The elastic barrier is
\begin{eqnarray}
F_e &=& 4\pi \int_{r_{cage}}^\infty dr r^2\rho(1-\alpha) g(r) \frac{K_0 u^2(r)}{2}\nonumber\\ 
&=& 12\Phi(1-\alpha) K_0\Delta r_{eff}^2\left(\frac{r_{cage}}{d}\right)^3\frac{1+\cfrac{\kappa r_{cage}}{2}}{(1 + \kappa r_{cage})^2},
\label{eq:16}
\end{eqnarray}
where the analytical expression in Eq.(\ref{eq:16}) adopts $g(r) \approx 1$ (relatively accurate for $r > r_{cage}$), and the factor ($1-\alpha$) arises since only mobile particles are responsible for elastic expansion. 

Based on Kramer's theory \cite{17,19}, the structural relaxation time is
\begin{eqnarray}
\frac{\tau_\alpha}{\tau_s} = 1+ \frac{2\pi}{\sqrt{K_0K_B}}\frac{k_BT}{d^2}\exp\left(\frac{F_B+F_e}{k_BT}\right),
\label{eq:tau}
\end{eqnarray}
where $K_B\equiv K_B(\alpha)=\left.\cfrac{\partial^2 F_{dyn}(r)}{\partial r^2}\right|_{r=r_B}$ is the barrier curvature and $\tau_s$ is a short relaxation time scale which is \cite{17}
\begin{eqnarray}
\tau_s = \frac{g(d)}{24\rho d^2}\sqrt{\frac{M}{\pi k_BT}}\left[1+\frac{1}{36\Phi}\int_{0}^{\infty}dq \frac{q^2\left[ S(q)-1 \right]^2}{S(q)+b(q)} \right], \nonumber\\
\label{eq:short}
\end{eqnarray}
where $b^{-1}(q) = 1-j_0(q)+2j_2(q)$, $j_n(x)$ is the spherical Bessel function of order $n$, $g(d)$ is a contact number among molecules, $M$ is the molar mass of particle, and $S(q)=1/\left[1-\rho c(q)\right]$. We assume that $\tau_s$ is unaffected by pinning. For water molecule, $d = 3$ $\AA$ and $M = 3\times 10^{-26}$ kg.

\section{Results and discussions}
\begin{figure}[htp]
\center
\includegraphics[width=9cm]{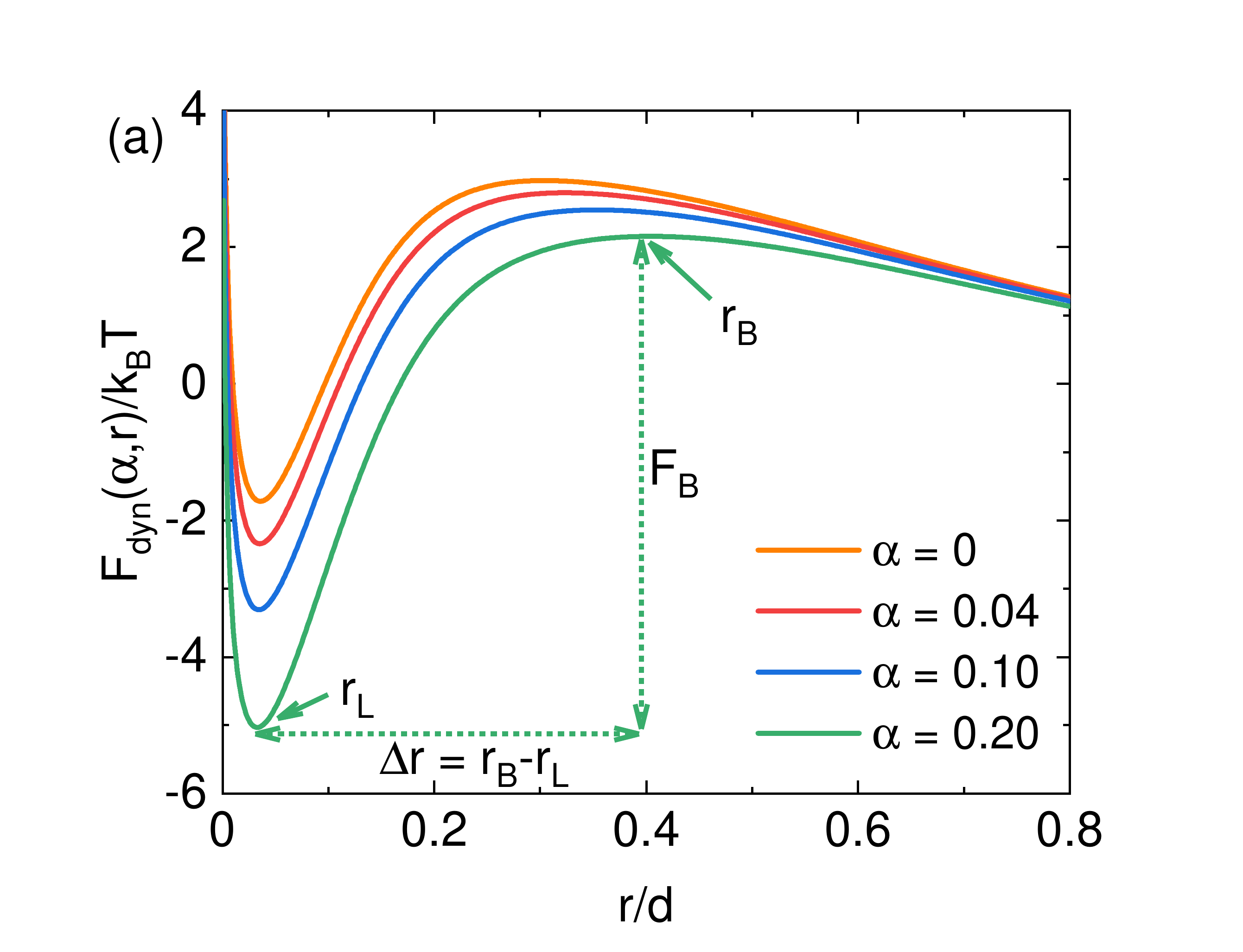}
\includegraphics[width=9cm]{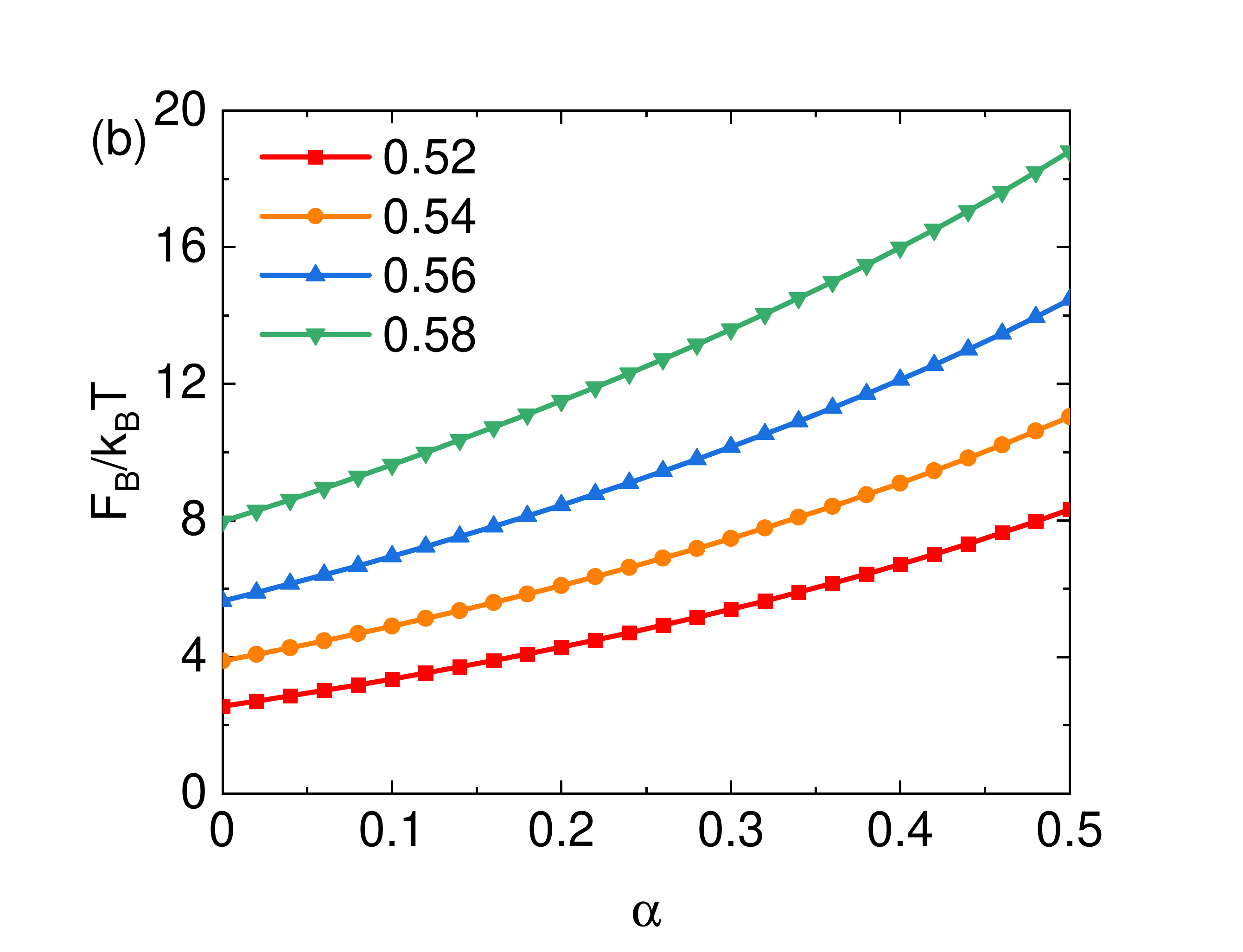}
\caption{\label{fig:3} (a) The dynamic free energy of a randomly pinned particle system with $\Phi=0.55$ calculated using Eq. (\ref{eq:1}) at several pinning fractions and (b) local barrier as a function of $\alpha$ at several volume fractions.}
\end{figure}

Figure \ref{fig:3}a shows a variation of the dynamic free energy with pinning. Although the structure of the pinned hard-sphere fluid remains unchanged with the pinning process, immobile particles strengthens the local caging constraint. Thus, the localization length decreases, and the barrier position and local barrier increase as $\alpha$ grows. Our results in Fig. \ref{fig:3}b suggest that the local barrier raises at least 2 times when $\alpha$ changes from 0 to 0.5. It means that pinning particles induces slower cage-scale dynamics compared to the unpinned system. 

\begin{figure}[htp]
\center
\includegraphics[width=9cm]{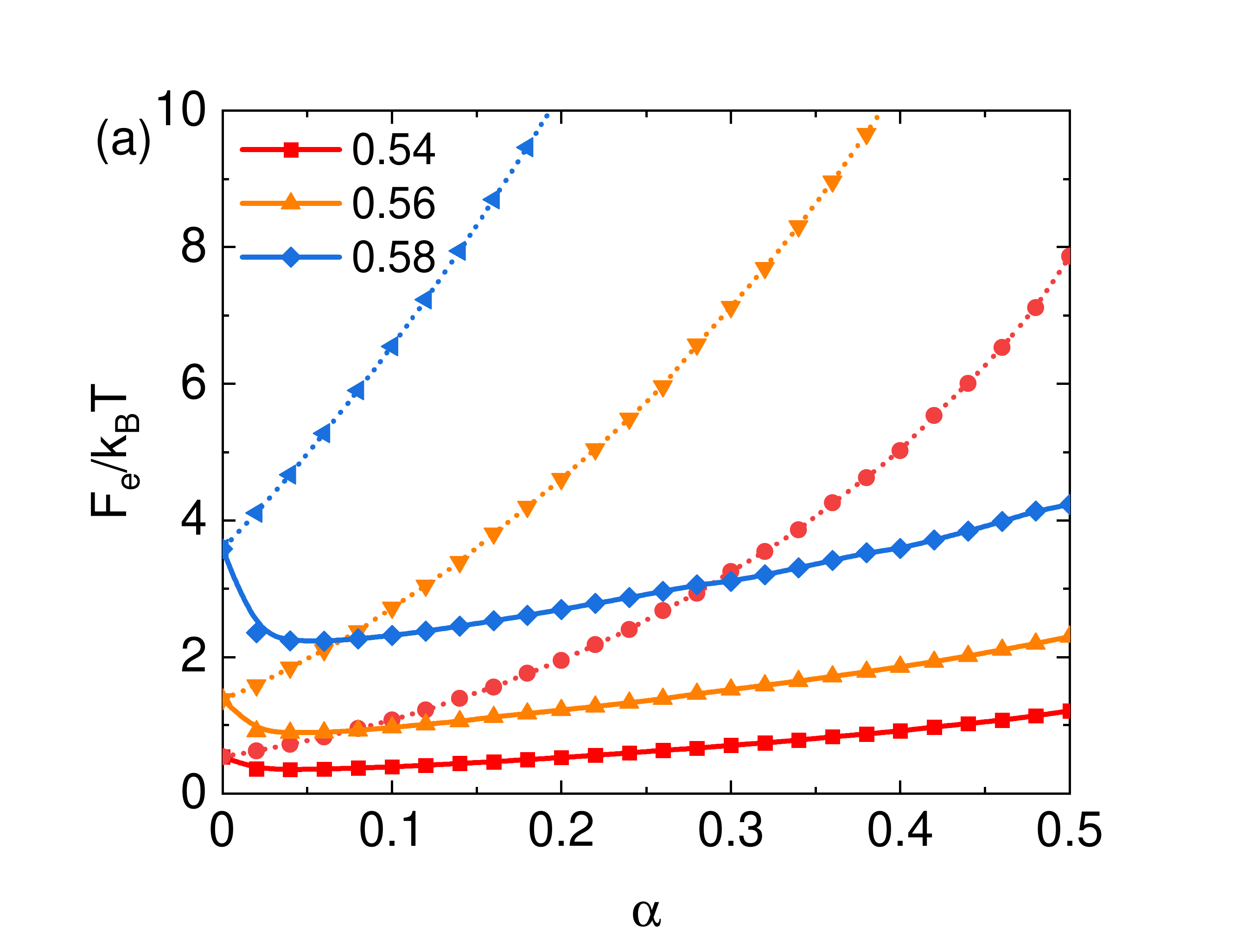}
\includegraphics[width=9cm]{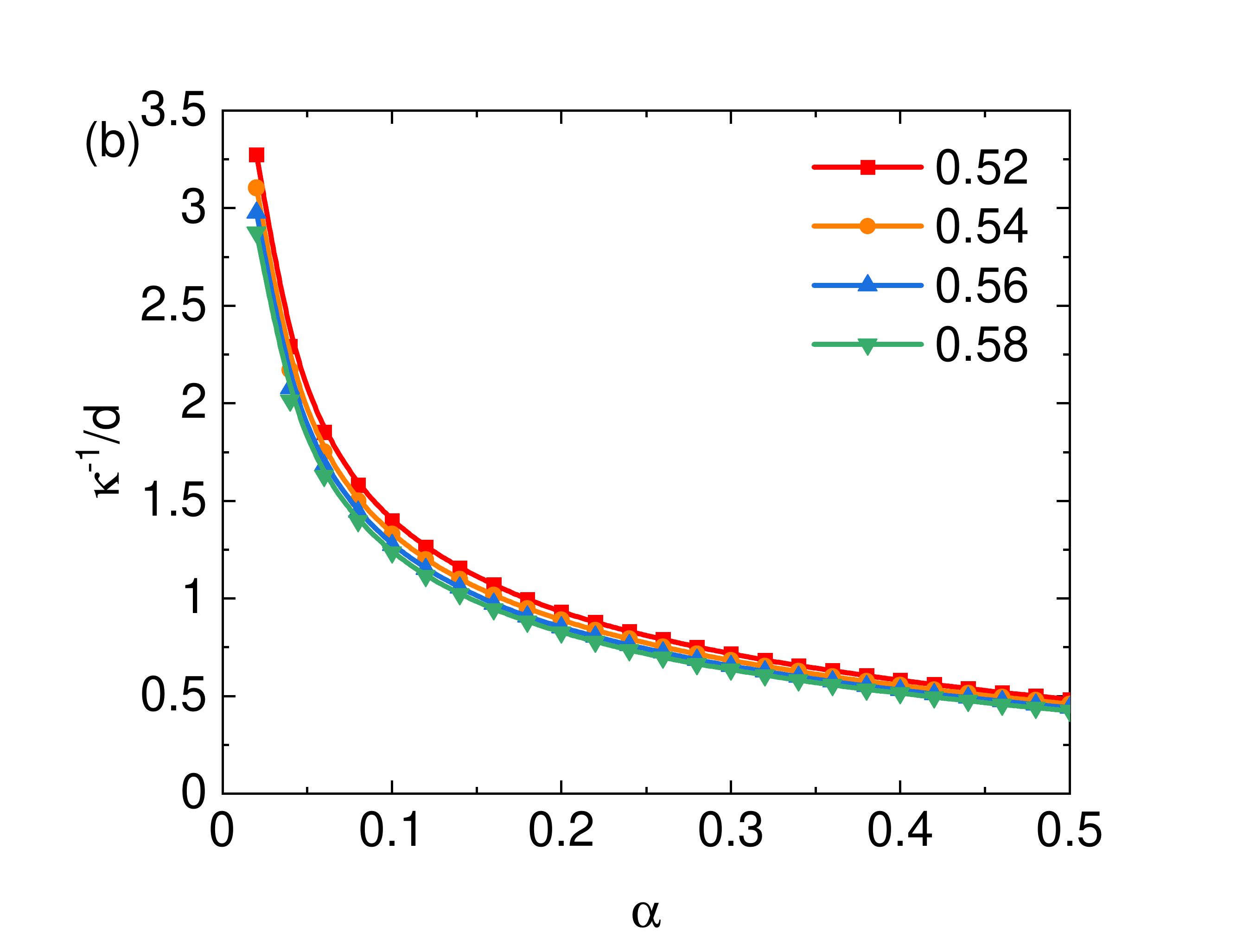}
\caption{\label{fig:2} (a) Screened (solid curves) and unscreened (dotted curves) elastic barrier as a function of $\alpha$ at various volume fractions. (b) The screening length as a function of $\alpha$ at various volume fractions.}
\end{figure}

Equation (\ref{eq:16}) and (\ref{eq:tau}) indicate how the alpha relaxation depends the elastic barrier. One can expect that the dynamics with and without the screening effect of elasticity deviate with each other by many orders of magnitude. When the screening effect is ignored in the same way as Ref. \cite{27}, $\kappa = 0$ and the analytical form of the elastic barrier becomes
\begin{eqnarray}
F_e = 12\Phi(1-\alpha) K_0\Delta r_{eff}^2\left(\frac{r_{cage}}{d}\right)^3.
\label{eq:16-1}
\end{eqnarray}
$ K_0(\alpha)\Delta r_{eff}^2(\alpha)\sim K_0(\alpha)\Delta r^4(\alpha)$ and the jump distance $\Delta r(\alpha)$ increase with $\alpha$. Although the factor ($1-\alpha$) is reduced as pinning more particles, the unscreened elastic barrier monotonically increases with $\alpha$ as shown in Fig. \ref{fig:2}a. This universal variation was already found in Ref. \cite{27}. 

When the screening effect is considered, $\kappa\neq 0$ and the displacement field decays with distance by both exponential function and inverse power laws of distance (as seen in Eq. (\ref{eq:dis})). The elasticity or collective effect on glassy dynamics is localized. The localization of elasticity is characterized by an exponential decay length, $\kappa^{-1}/d$. Figure \ref{fig:2}b shows a weak dependence of the decay length on the volume fraction. Increasing the number of pinned particles leads to greater screening effect of the long range cooperative rearrangement, shortens $\kappa^{-1}$, and significantly decreases $F_e$. At the same pinning fraction, the screened $F_e$ calculated using Eq. (\ref{eq:16}) is much smaller than the unscreened counterpart. Interestingly, the screened $F_e$ non-monotonically varies with $\alpha$. After a considerable drop at $\alpha \leq 0.06$, the elastic energy has a small increase but it cannot compare to a rise of the local barrier.

Note that the alpha time calculated using Eqs. (\ref{eq:tau}) and (\ref{eq:short}) is density-dependent. To validate our screened ECNLE theory, it is necessary to compare the temperature dependence of the predicted alpha time of water with simulation. The comparison requires a thermal mapping to convert from an effective hard-sphere volume fraction into temperature. In prior works \cite{Phan2019,PhanPRL,35,40,41,42}, we formulated a universal correlation between density and temperature for various amorphous materials based on thermal expansion process. The thermal mapping is 
\begin{eqnarray}
T = T_{g,bulk}+\frac{\Phi_g-\Phi}{\beta\Phi_0},
\label{eq:thermalmapping}
\end{eqnarray}
where $T_{g,bulk}$ and $\Phi_{g}$ are the bulk glass transition temperature and the glass transition volume fraction, respectively, defined by $\tau_\alpha(\Phi_g)=\tau_\alpha(T_{g,bulk})=100$s, $\beta=12\times10^{-4}$ $K^{-1}$ is an effective thermal expansion coefficient, and $\Phi_0=0.5$ is a characteristic volume fraction \cite{Phan2019,PhanPRL,35,40,41,42}. Equation (\ref{eq:thermalmapping}) is a minimalist conversion and all chemical/conformational complexities of molecules are encoded in $T_{g,bulk}$.

\begin{figure}[htp]
\center
\includegraphics[width=9cm]{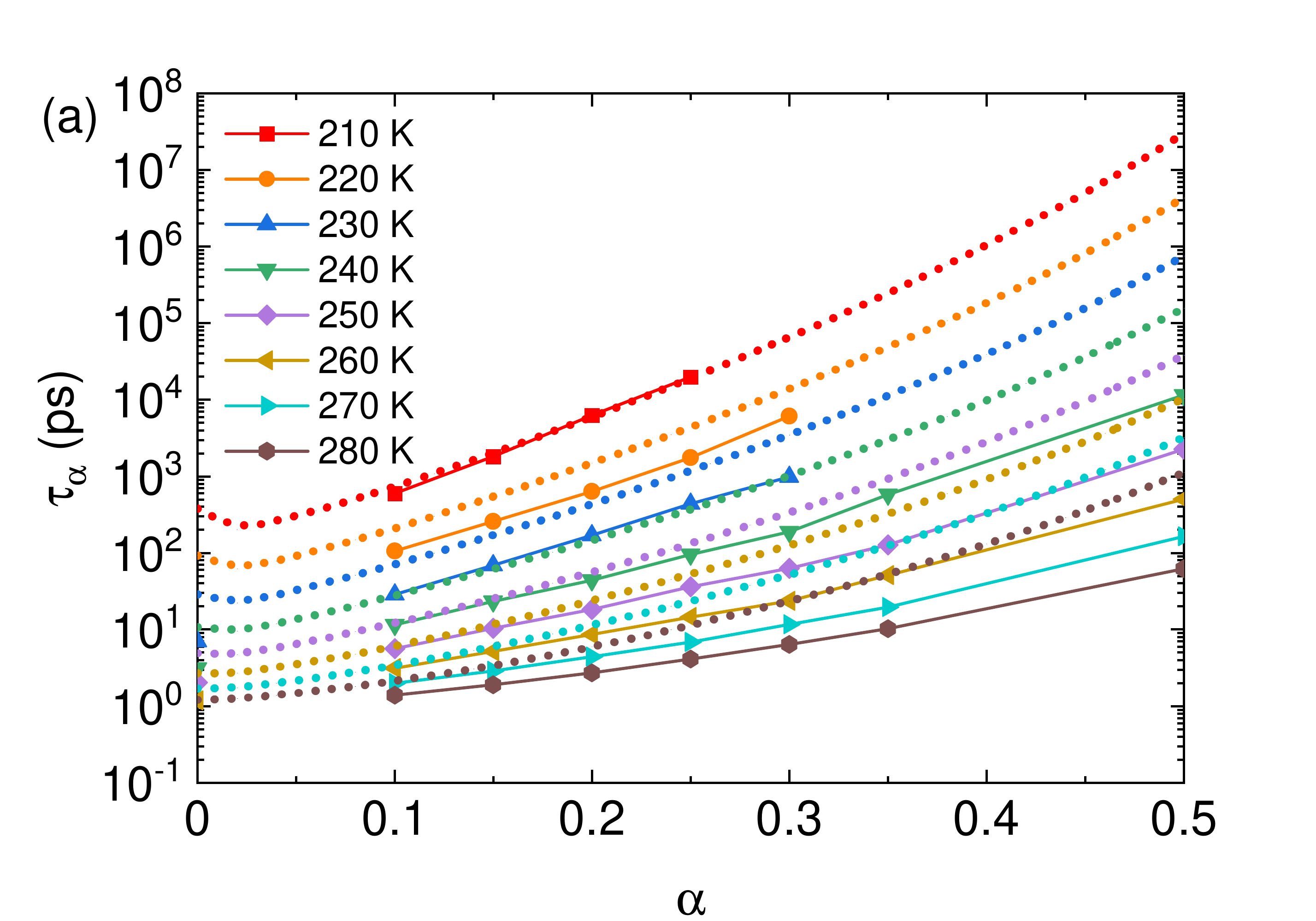}
\includegraphics[width=9cm]{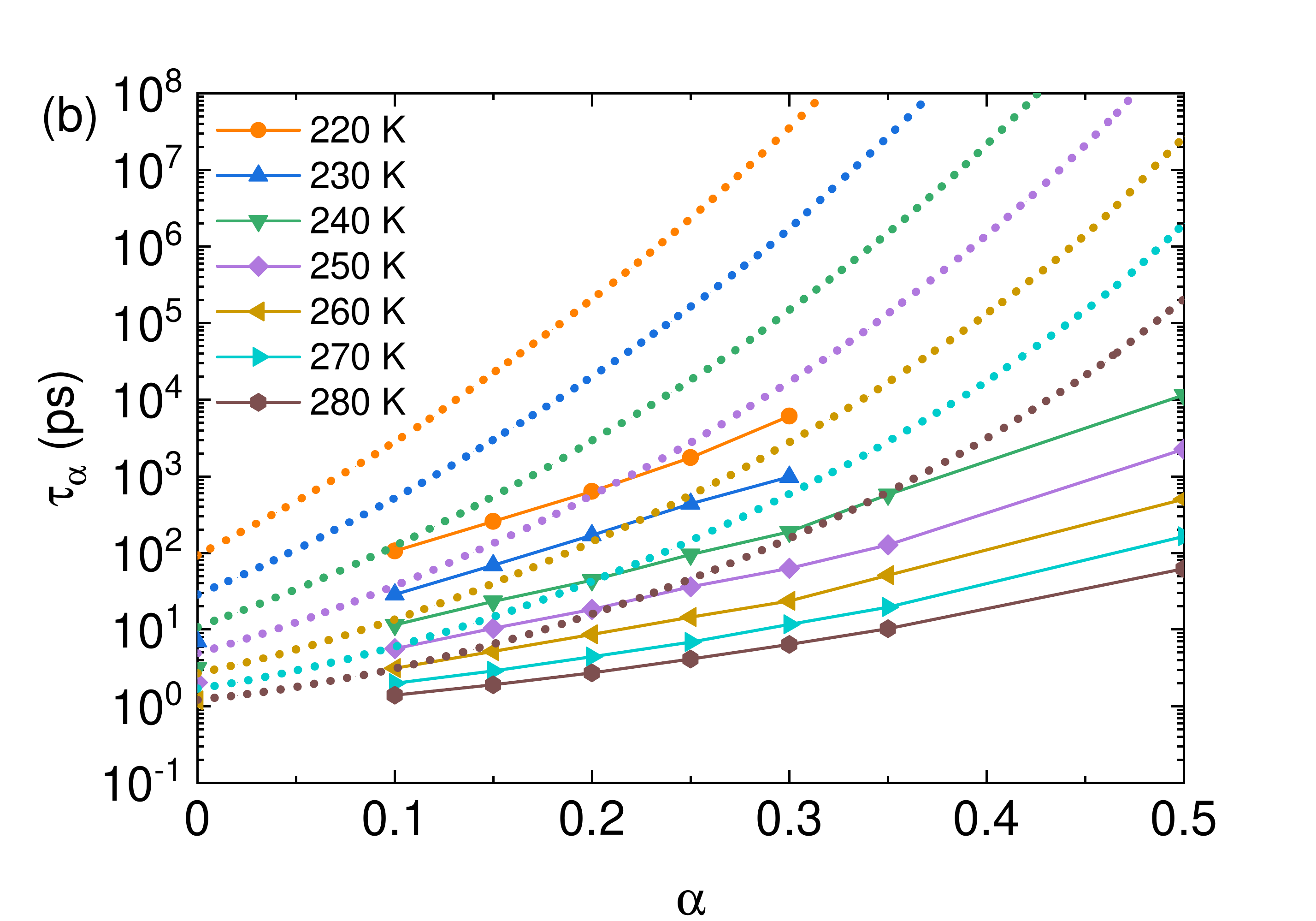}
\caption{\label{fig:4} The alpha time in picoseconds calculated (a) with and (b) without the displacement field screening (dotted curves) as a function of $\alpha$ compared to simulation results \cite{7,8} (points with solid curves) for water at various temperatures in Kelvin.}
\end{figure}

For unpinned bulk water, we find $\Phi_g=0.6191$ and $T_{g,bulk}\approx 136$ $K$ is the most commonly accepted experimental value \cite{33,34,36,37}.  We now can use the ENCLE theory to calculate $\tau_\alpha(T)$ for pinned-mobile water and compare with thermal simulations \cite{7,8} in Fig. \ref{fig:4}a. The screened ECNLE approach provides a good description with simulation. Particularly, the predicted relaxation time for $T= 210$ $K$ perfectly overlaps the simulated relaxation time over a wide range of pinning fractions. However, quantitative agreement between theory and simulation are not obtained at all temperatures. A possible reason is that we assume a universal coupling between the cage-scale hopping and molecular cooperative rearrangements in our calculations, thus relatively simplifying geometrical and chemical effects. This treatment allows us to predict $\tau_\alpha(T)$ without any adjustable parameter. In Ref. \cite{Phan2019,46}, one material-specific numerical parameter $a_c$ is introduced to scale the collective elastic barrier as $F_e \rightarrow a_cF_e$. The parameter characterizes for a non-universal local-collective correlation in polymers and other amorphous materials. By adjusting $a_c$, the best fit to experimental $\tau_\alpha(T)$ and dynamic fragility are achieved. But when experimental/simulation data is not available, the parameter $a_c$ cannot be determined and analysis is not predictive.

Ignoring the screening effect leads to poor agreement with simulation as shown in Fig. \ref{fig:4}b since $\tau_\alpha$ grows more rapidly with $\alpha$. Particularly for $\alpha \geq 0.2$, a separation between $\tau_\alpha$ computed using simulation and the ECNLE theory without screening becomes remarkable. For $\alpha \leq 0.1$, $\tau_\alpha$ in Fig. \ref{fig:4}a has a slight drop because of a decrease of the elastic barrier but substantially increases when $\alpha \geq 0.1$. Since $\tau_\alpha(T)$ non-monotonically varies with pinning, one can expect the same behavior for the $T_g$ variation. It is hard to know whether the former trend is correct because the simulations does not study the low $\alpha$ regime. The reduction may be a true nature of the phenomenon or an artifact of our mean field like screening analysis at low $\alpha$.

\begin{figure}[htp]
\center
\includegraphics[width=9cm]{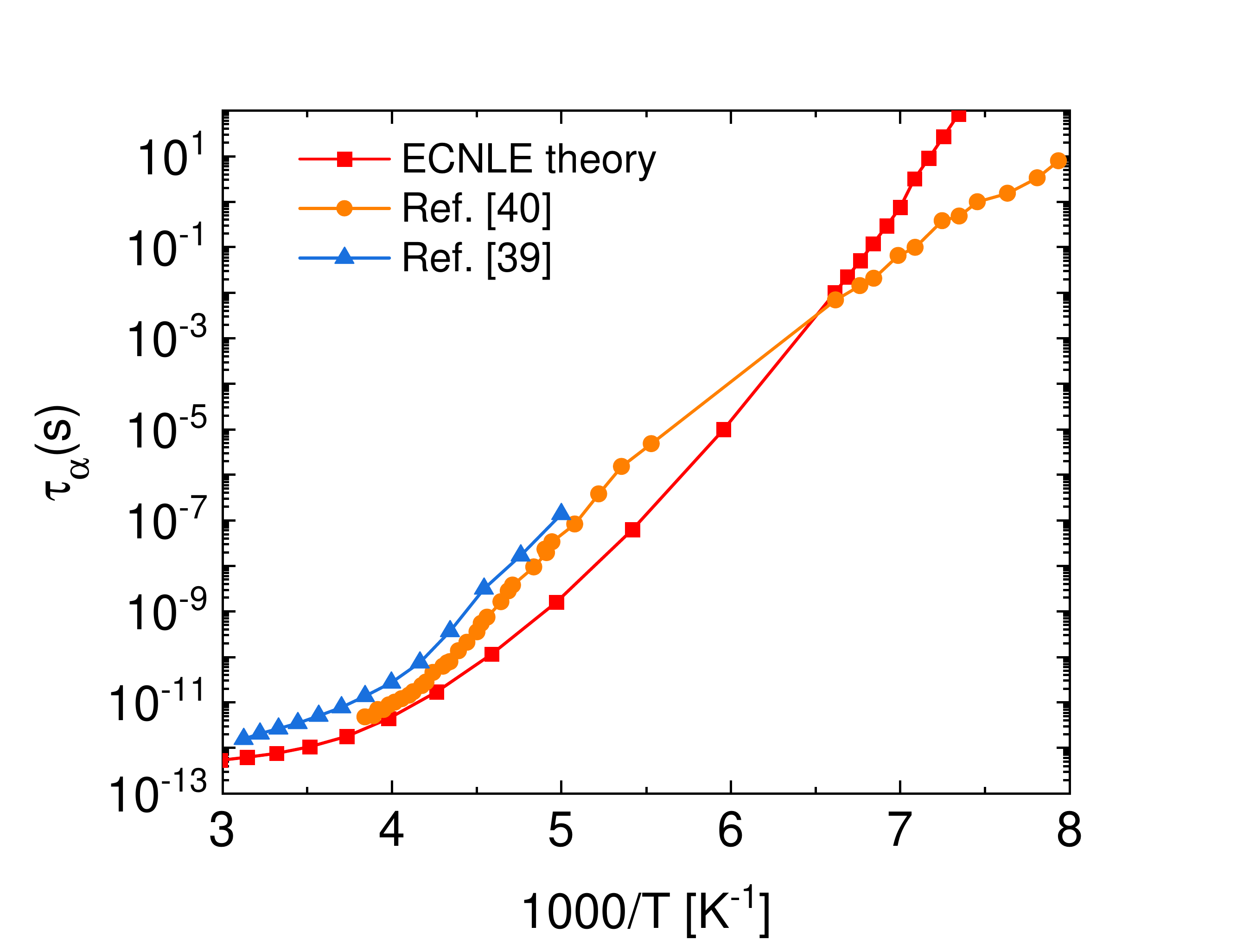}
\caption{\label{fig:5-1} The temperature dependence of the alpha time of water without pinning.}
\end{figure}

Equations (\ref{eq:tau}) and (\ref{eq:thermalmapping}) can be also exploited to determine the temperature dependence of $\tau_\alpha$ for unpinned bulk water and contrast numerical results with a prior simulation \cite{38} and experimental study \cite{39}. In Ref. \cite{39}, authors measure the self diffusion constant in transiently heated water films as a function of temperature, $D(T)$. In the framework of ECNLE theory \cite{40,41,42}, the relation between the structural relaxation time and diffusion constant is $D(T)=\Delta r^2/6\tau_\alpha(T)$. Thus, by approximating $\Delta r\approx 0.5d = 1.5$ $\AA$, it is possible for us to deduce $\tau_\alpha(T)$ from diffusion data. Figure \ref{fig:5-1} shows that our predicted $\tau_\alpha$ agrees qualitatively well with experimental data for $T > 252$ $K$. This is because at high temperatures, the elastic barrier slightly contributes to the relaxation and the liquid-air interface has a minor effect on the diffusion measurement \cite{22,24}. Thus, the film-averaged diffusion is approximately identical to the bulk value. At low temperatures, interfacial effects on the diffusion and relaxation process become larger \cite{22,24}. The diffusion constant measured in film systems is quantitatively different from the bulk diffusion. While the temperature-dependent diffusion obtained using TIP4P/ICE model \cite{38} nearly overlaps our theoretical curve when reduced by a factor of $\sim 2$. Deviation between these curves becomes significant at low temperatures since the ECNLE theory does not capture physics of possible emergence of  crystallization and fragile-to-strong transition reported in Ref. \cite{37,43}.

The thermal dependence of $\tau_\alpha$ is also analyzed using the dynamic fragility $m=\left(\partial \log\tau_\alpha/\partial(T_g/T)\right)|_{T=T_{g}}$. From ECNLE data in Fig. \ref{fig:5-1}, we find that the theoretical fragility is about 46, which is much higher than the experimental analog ($m\sim 18$) \cite{39}. This difference can be considered as a consequence of the super-Arrhenius dynamics. Both the ECNLE theory and Dyre's phenomenological elastic shoving model \cite{44,45} indicate that the strong temperature dependence of the elastic collective dynamics with cooling is solely responsible for the super-Arrhenius behavior. However, in Ref. \cite{43}, Tanaka and his coworkers proposed a two-state theory to describe the water dynamics. Two states (so-called fast and slow water state) co-exist in different structures and obey Arrhenius law. The view contradicts with our initial assumption that only one structure exists at a given temperature/density. This may be another missing physics for the ECNLE theory to predict the fragile-to-strong crossover besides nonuniversal coupling of cage scale hopping and collective elastic distortion. 

\section{Conclusions}
We have formulated a new theoretical approach based on the ECNLE theory to understand physical mechanisms underlying slow dynamics in randomly pinned particles. The random pinning constructs neutral confinement which has no influence on pair structure but the dynamics is dramatically changed. Our theoretical results show that pinning particles induces slower cage-scale dynamics via strengthening a local caging constraint on mobile particles. An expansion of the first coordination shell is required to create space for a large amplitude local structural relaxation and generate a long range displacement field outside the cage via elastically cooperative motions of particles. Since the pinned particles are not allowed to move, the displacement field nucleated from a cage surface cannot propagate through them. Thus, we propose a new screening type of effect for the elastic displacement field and derive a mathematical form of screened displacement field and decay length. Increasing fraction of pinned particles localizes more collective motions. This screening effect leads to non-monotonic variation of both the elastic barrier and structural relaxation time with pinning at high densities. Although numerical results agree rather well with simulations \cite{7,8}, a slight drop of $\tau_\alpha$ at small pinning fraction needs more simulation studies to verify. For unpinned bulk water, our predicted $\tau_\alpha(T)$ is quantitatively consistent with experimental data in Ref. \cite{39} and simulation in Ref. \cite{38}. The agreements clearly validate the screened ECNLE theory and suggest that the screening effect of displacement field proposed in this work is relatively sensible. The approach could reveal a true nature of slow dynamics in randomly pinned particle fluids.

\begin{acknowledgments}
This research was funded by the Vietnam National Foundation for Science and Technology Development (NAFOSTED) under grant number 103.01-2019.318. 
\end{acknowledgments}
\newpage

\begin{thebibliography}{5}
\bibitem{1} P. G. Debenedetti and F. H. Stillinger, Nature {\bf 410}, 259-267 (2001).
\bibitem{2} K. Ngai, J. Non-Cryst. Solids {\bf 353}, 709–718 (2007).
\bibitem{3} L. Berthier and G. Biroli, Rev. Mod. Phys. {\bf 83}, 587 (2011).
\bibitem{4} F. Kremer, \emph{Dynamics in Geometrical Confinement}, Springer, Berlin, 2014.
\bibitem{5} K. S. Schweizer and D. S. Simmons, J. Chem. Phys. {\bf 151}, 240901 (2019).
\bibitem{6} S. Gokhale,	K. Hima Nagamanasa,	R. Ganapathy, and A. K. Sood, Nature Commun. {\bf 5}, 4685 (2014).
\bibitem{7} F. Klameth and M. Vogel, J. Phys. Chem. Lett. {\bf 6}, 4385 (2015).
\bibitem{8} F. Klameth, P. Henritzi, and M. Vogel, J. Chem. Phys. {\bf 140}, 144501 (2014).
\bibitem{29} W. Kob and L. Berthier, Phys. Rev. Lett. {\bf 110}, 245702 (2013).
\bibitem{30} P. Scheidler, W. Kob, and K. Binder, J. Phys. Chem. B {\bf 108}, 6673-6686 (2004).
\bibitem{31} W. Kob and D. Coslovich, Phys. Rev. E {\bf 90}, 052305 (2014).
\bibitem{9} R. B. Jadrich and K. S. Schweizer, Phys. Rev. Lett. {\bf 113}, 208302 (2014).
\bibitem{10} P. S. Sarangapani, A. B. Schofield, and Y. Zhu, Phys. Rev. E {\bf 83}, 030502 (2011).
\bibitem{11} Z. Fakhraai and J. A. Forrest, Phys. Rev. Lett. {\bf 95}, 025701 (2006).
\bibitem{12} E. C. Glor, R. J. Composto, and Z. Fakhraai, Macromolecules {\bf 48}, 6682 (2015).
\bibitem{13} R. J. Lang, W. L. Merling, and D. S. Simmons, ACS Macro Lett. {\bf 3}, 758 (2014).
\bibitem{14} K. Paeng and M. D. Ediger, Macromolecules {\bf 44}, 7034 (2011).
\bibitem{15} K. Paeng, R. Richert, and M. D. Ediger, Soft Matter {\bf 8}, 819–826 (2012).
\bibitem{16} S. Mirigian and K. S. Schweizer, J. Chem. Phys. {\bf 141}, 161103 (2014).
\bibitem{17} S. Mirigian and K. S. Schweizer, J. Chem. Phys. {\bf 140}, 194506 (2014).
\bibitem{27} A. D. Phan and K. S. Schweizer, J. Chem. Phys. {\bf 148}, 054502 (2018).
\bibitem{Phan2019} A. D. Phan, K. Wakabayashi, M. Paluch, V. D. Lam, RSC Adv. \textbf{9}, 40214-40221 (2019). 
\bibitem{PhanPRL} A. D. Phan, A. Zaccone, V. D. Lam, and K. Wakabayashi, Phys. Rev. Lett. \textbf{126}, 025502 (2021)
\bibitem{35} A. D. Phan, J. Knapik-Kowalczuk, M. Paluch, T. X. Hoang, and K. Wakabayashi, Mol. Pharm. {\bf 16}, 2992-2998, (2019).
\bibitem{40} A. D. Phan, K. Koperwas, M. Paluch, and
K. Wakabayashi, Phys. Chem. Chem. Phys. {\bf 22}, 24365 (2020).
\bibitem{41} N. K. Ngan, A. D. Phan, and A. Zaccone, Phys. Status Solidi RRL {\bf 15}, 2100235 (2021).
\bibitem{42} A. D. Phan, N. K. Ngan, D. T. Nga, N. B. Le, and C. V. Ha, Phys. Status Solidi RRL {\bf 16}, 2100496 (2022).
\bibitem{18} R. Zhang and K. S. Schweizer, J. Phys. Chem. B {\bf 122}, 3465−3479 (2018).
\bibitem{21} A. D. Phan and K. S. Schweizer, Macromolecules {\bf 52}, 5192-5206 (2019).
\bibitem{24} A. D. Phan and K. S. Schweizer, Macromolecules {\bf 51}, 6063-6075 (2018).
\bibitem{22} A. Ghanekarade, A. D. Phan, Kenneth S. Schweizer, and D. S. Simmons, PNAS {\bf 118}, e2104398118 (2021).
\bibitem{20} D. C. Viehman and K. S. Schweizer, J. Phys. Chem. {\bf 128}, 084509 (2007).
\bibitem{19} S. Mirigian and K. S. Schweizer, J. Chem. Phys. {\bf 140}, 194507 (2014).

\bibitem{23} K. S. Schweizer and G. Yatsenko, J. Chem. Phys. {\bf 127}, 164505 (2007).
\bibitem{33} D. Bertolini, M. Cassettari, G. Salvetti, J. Chem. Phys. {\bf 76}, 3285-3290 (1982).
\bibitem{34} L. Kringle, W. A. Thornley, B. D. Kaya, and G. A. Kimmel, PNAS {\bf 118}, e2022884118 (2021).

\bibitem{36} K. Amann-Winkel, C. Gainaru, P. H. Handle, M. Seidl, H. Nelson, R. Bohmer, and T. Loerting, PNAS {\bf 110}, 17720–17725  (2013).
\bibitem{37} J. Swenson, Phys. Chem. Chem. Phys. {\bf 20}, 30095 (2018).

\bibitem{46} S.-J. Xie and K. S. Schweizer, Macromolecules {\bf 49}, 9655-9664 (2016).




\bibitem{38} J. R. Espinosa, C. Navarro, E. Sanz, C. Valeriani, and C. Vega, J. Chem. Phys. {\bf 145}, 211922 (2016).
\bibitem{39} Y. Xu, N. G. Petrik, R. S. Smith, B. D. Kay, and G. A. Kimmel, PNAS {\bf 113}, 14921-14925 (2016).
\bibitem{43} R. Shi, J. Russo, and H. Tanaka, PNAS {\bf 115}, 9444-9449 (2018).
\bibitem{44} J. C. Dyre, T. Christensen, and N. B. Olsen, J. Non-Cryst. Solids {\bf 352}, 4635 (2006). 
\bibitem{45} J. C. Dyre, Rev. Mod. Phys. {\bf 78}, 953 (2006).

\end{thebibliography}

\end{document}